\documentclass[12pt]{article}
 \usepackage{amssymb}
 \usepackage{latexsym}

\renewcommand{\theequation}{\arabic{section}.\arabic{equation}}
\catcode`\@=11 \@addtoreset{equation}{section}\catcode`\@=12

\def\barr{\left(\begin{array}}
\def\earr{\end{array}\right)}

\newcommand{\R}{ {\mathbb R} }

\newcommand{\fnm}{\footnotemark}
\newcommand{\fnt}{\footnotetext}

\begin{document}

\begin{center}

\large \bf On anisotropic Gauss-Bonnet cosmologies in (n+1)
dimensions, governed by an n-dimensional Finslerian 4-metric

\end{center}

\vspace{0.3truecm}

\begin{center}

\normalsize\bf V. D. Ivashchuk\fnm[1]\fnt[1]{e-mail:
ivashchuk@mail.ru},

\vspace{0.3truecm}

\it Center for Gravitation and Fundamental Metrology, VNIIMS, 46
Ozyornaya ul., Moscow 119361, Russia

\it Institute of Gravitation and Cosmology, Peoples' Friendship
University of Russia,  6 Miklukho-Maklaya ul., Moscow 117198,
Russia

\end{center}

\begin{abstract}

The $(n +1)$-dimensional  Einstein-Gauss-Bonnet (EGB) model is
considered. For  diagonal  cosmological  metrics, the equations of
motion are written as a set of Lagrange equations   with the
effective  Lagrangian   containing  two ``minisuperspace'' metrics
on   $\R^{n}$: a 2-metric  of  pseudo-Euclidean signature  and  a
Finslerian  4-metric  proportional to the  $n$-dimensional
Berwald-Moor 4-metric. For the case of the ``pure'' Gauss-Bonnet
model, two exact solutions are presented, those with power-law and
exponential dependences  of the scale factors (w.r.t. the
synchronous time variable). (The power-law solution was considered
earlier by N. Deruelle,  A. Toporensky, P. Tretyakov, and S.
Pavluchenko.)   In the case of EGB cosmology, it is shown that for
any non-trivial  solution with an exponential dependence of scale
factors, $a_i(\tau) = A_i \exp( v^i \tau)$, there are no more than
three different  numbers among $v^1,...,v^n$.

\end{abstract}

%\vspace{1.3truecm}

%{\it Key words:} multidimensional cosmology, anisotropic
%solutions, Einstein-Gauss-Bonnet gravity, Finslerian Berwald-Moor
%metric.

%\vspace{0.3truecm}

%{\it PACS numbers:} 04.50.-h, 04.50.Kd, 04.60.Cf, 98.80.Jk,
%11.10.Ef

\newpage

%%%%%%%%%%%%%%%%%%%%%%%%%%%%%%%%%%%%%%%%%%%
\section{Introduction}
%%%%%%%%%%%%%%%%%%%%%%%%%%%%%%%%%%%%%%%%%%

In this paper, we consider $D$-dimensional gravitational model
with the Gauss-Bonnet term. The action reads
\begin{equation}
S =  \int_{M} d^{D}z \sqrt{|g|} \{ \alpha_1 R[g] + \alpha_2 {\cal
L}_2[g] \}, \label{1.1}
\end{equation}

where $g = g_{MN} dz^{M} \otimes dz^{N}$ is a metric defined on
the manifold $M$, ${\dim M} = D$, $|g| = |\det (g_{MN})|$, and

\begin{equation}
{\cal L}_2 = R_{MNPQ} R^{MNPQ} - 4 R_{MN} R^{MN} +R^2 \label{1.2}
\end{equation}
is the standard Gauss-Bonnet term. Here $\alpha_1$ and $\alpha_2$
are constants.  The appearance of the  Gauss-Bonnet term in
multidimensional gravity is motivated by string theory
\cite{GB-strings,Zwiebach}.

At present, the so-called Einstein-Gauss-Bonnet (EGB) gravity and
its modifications are intensively used in  cosmology, see
\cite{NojOd0,CElOdZ} (for $D =4$), \cite{Ishihara}-\cite{KirMak}
and references therein,  e.g., for explanation  of the accelerated
expansion of the Universe following from the supernovae (type Ia)
observational data \cite{Kowalski}. Certain exact solutions in
multidimesional EGB cosmology  were obtained in
\cite{Ishihara}-\cite{KirMak} and some other papers.

EGB gravity is also  intensively investigated  in the context of
black-hole physics. The  most  important results here are related
to  the well-known Boulware-Deser-Wheeler solution
\cite{BoulDesWheel} and its generalizations \cite{Wheel-general},
for a review  and references see \cite{GB-bh-rev}. (For certain
applications of  brane-world models with the Gauss-Bonnet term see
also the review \cite{Br} and references therein.)

Here we are interested in  cosmological solutions with diagonal
metrics,  governed by time-dependent scale factors.

For  $\alpha_2 = 0$ we have  the Kasner-type solution with the
metric
\begin{equation}
g= - d \tau \otimes d \tau  + \sum_{i=1}^{n}  A_i^2 \tau^{2p^i}
dy^i \otimes dy^i, \label{1.3}
\end{equation}
where  $A_i > 0$ are arbitrary constants, $D = n +1$,   and
parameters  $p^i$ obey the relations $\sum_{i=1}^n  p^i =
\sum_{i=1}^n  (p^i)^2  = 1$  and hence $ \sum_{ 1 \leq i < j \leq
n} p^i p^j =  0$.  For $D =4$ it is the well-known Kasner solution
\cite{Kasner}.

In \cite{Deruelle}, the Einstein-Gauss-Bonnet (EGB) cosmological
model  was  considered (see also \cite{Lanc}). For  the ``pure''
Gauss-Bonnet (GB) case $\alpha_1 = 0$ and  $\alpha_2 \neq 0$, N.
Deruelle has obtained a cosmological  solution with the metric
(\ref{1.3}) for $n = 4, 5$  and parameters obeying  the relations
\begin{equation}
\sum_{i=1}^n  p^i = 3,  \qquad \sum_{1 \leq i < j < k < l \leq n}
p^i p^j p^k p^l = 0  \label{1.7}.
\end{equation}
It was  reported by A. Toporensky and P. Tretyakov in \cite{TT}
that this solution was  verified  by them for $n = 6,7$. In the
recent paper by S. Pavluchenko \cite{Pavl},  the power-law
solution was verified for all $n$ (and also generalized   to the
Lowelock case \cite{Low}).

In this paper  we give a  derivation of  the ``power-law''
solution (\ref{1.3}),  (\ref{1.7}) and a solution with the
exponential dependence of scale factors  for arbitrary $n$. We
note that the recent numerical analysis  of cosmological solutions
in EGB gravity for $D =  5, 6$ \cite{PTop}  shows  that the
singular  solution    (\ref{1.3}), (\ref{1.7}) (e.g. with a little
generalization of the scale factors  $a_i(\tau) = A_i (\tau_0 \pm
\tau)^{p^i}$, where $\tau_0$ is constant)  may appear as an
asymptotical solution for certain  initial values as well as the
Kasner-type solution  does.

The paper is organized as follows. In Section 2, the equations of
motion for $(n +1)$-dimensional EGB model are considered.   For
diagonal  cosmological  metrics,  the equations of motion are
written in the form of  Lagrange equations  corresponding to a
certain ``effective'' Lagrangian  (see also \cite{Deruelle,Pavl}).
Section 3  deals with  the ``pure'' Gauss-Bonnet model. Here two
exact solutions are obtained, with power-law and exponential
dependences of the scale factors on the synchronous time variable.
In Section 4, the equations of motion are reduced to an autonomous
set of first order differential equations in terms of synchronous
time variable $\tau$.   For $\alpha_1 \neq 0$ and $\alpha_2 \neq
0$,   we show that for any non-trivial solution with the
exponential dependence of the scale factors $a_i(\tau) = A_i \exp(
v^i \tau)$, $i = 1,...,n$,   there are no more than three
different numbers among  $v^1,...,v^n$.

%%%%%%%%%%%%%%%%%%%%%%%%%%%%%%%%%%%%%%%%%%%%%%%%%%%%%%%%%%%%%%%%%%
\section{The cosmological  model and its effective Lagrangian}
%%%%%%%%%%%%%%%%%%%%%%%%%%%%%%%%%%%%%%%%%%%%%%%%%%%%%%%%%%%%%%%%%%

\subsection{The set-up }

We consider the manifold
\begin{equation}
M = \R_{*}  \times M_{1} \times \ldots \times M_{n}, \label{2.1}
\end{equation}
with the metric
\begin{equation}
g= - e^{2{\gamma}(t)} dt \otimes dt  + \sum_{i=1}^{n}
e^{2\beta^i(t)}  dy^i \otimes dy^i, \label{2.2}
\end{equation}
where $M_i$ is a 1-dimensional manifold with the metric $g^i =
dy^i \otimes dy^i$,  $i = 1, \dots, n$. Here and henceforth,
$\R_{*} = (t_{-},t_{+})$ is an open subset in $\R$. (The functions
${\gamma}(t)$ and  $\beta^i(t)$,  $i = 1,\ldots, n$, are smooth on
$\R_{*}$.)

The integrand in  (\ref{1.1}), if the metric (\ref{2.2}) is
substituted, reads  (see  \cite{Deruelle} for $\gamma = 0$)

\begin{equation}
\sqrt{|g|} \{ \alpha_1 R[g] + \alpha_2  {\cal L}_2[g] \} = L +
\frac{df}{dt},   \label{2.3}
\end{equation}
where
\begin{eqnarray}
L = \alpha_1 e^{-\gamma + \gamma_0} G_{ij} \dot{\beta}^i
\dot{\beta}^j - \frac{1}{3} \alpha_2 e^{- 3 \gamma + \gamma_0}
G_{ijkl} \dot{\beta}^i \dot{\beta}^j \dot{\beta}^k \dot{\beta}^l,
\label{2.4}
\end{eqnarray}
$\gamma_0 = \sum_{i =1}^{n} \beta^i$
and
\begin{eqnarray}
G_{ij} = \delta_{ij} -1,
\label{2.10}   \\
G_{ijkl}  = (\delta_{ij} -1)(\delta_{ik} -1)(\delta_{il} -1)
(\delta_{jk} -1)(\delta_{jl} -1)(\delta_{kl} -1) \label{2.11}
\end{eqnarray}
are  the components of two ``minisuperspace'' metrics on $\R^{n}$.
The first one is the well-known 2-metric of pseudo-Euclidean
signature: $<v_1,v_2> = G_{ij}v^i_1 v^j_2$, while second one is
the Finslerian 4-metric: $<v_1,v_2,v_3,v_4> = G_{ijkl}v^i_1 v^j_2
v^k_3 v^l_4$,  $v_s = (v^i_s) \in \R^n$, where $<.,.>$ and
$<.,.,.,.>$ are respectively    $2$- and $4$-linear symmetric
forms on $\R^n$. (We denote  $\dot{A} = dA/dt$.)

The  function $f = f(\gamma, \beta, \dot{\beta})$  in (\ref{2.3})
is  presented in  Appendix B.

The derivation of (\ref{2.4}) is based on the  relations from
Appendix A and  the following identities
\begin{eqnarray}
G_{ij}v^i v^j = \sum_{i =1}^{n} (v^i)^2 - (\sum_{i =1}^{n} v^i )^2,
\label{2.12}   \\
G_{ijkl}v^i v^j v^k v^l  = (\sum_{i =1}^{n} v^i )^4 - 6 (\sum_{i
=1}^{n} v^i )^2  \sum_{j =1}^{n} (v^j)^2
\nonumber \\
+ 3 ( \sum_{i =1}^{n} (v^i)^2 )^2  + 8  (\sum_{i =1}^{n} v^i )
\sum_{j =1}^{n} (v^j)^3  - 6 \sum_{i =1}^{n} (v^i)^4.
\label{2.13}
\end{eqnarray}

It immediately follows  from the definitions (\ref{2.10}) and
(\ref{2.11}) that
\begin{eqnarray}
G_{ij}v^i v^j = -2 \sum_{i < j} v^i v^j,
\label{2.14}   \\
G_{ijkl}v^i v^j v^k v^l  = 24 \sum_{i < j < k < l} v^i v^j v^k v^l
. \label{2.15}
\end{eqnarray}

Due to (\ref{2.15})  $G_{ijkl}v^i v^j v^k v^l$ is zero for $n = 1,
2, 3$ ($D = 2, 3, 4$).  For $n = 4$ ($D = 5$),  $G_{ijkl}v^i v^j
v^k v^l = 24 v^1 v^2 v^3 v^4$ and our  4-metric is proportional to
the well-known Berwald-Moor 4-metric \cite{Berwald,Moor} (see also
\cite{AsanPon,Bogos,GarPav} and references therein). We remind the
reader that  the 4-dimensional Berwald-Moor 4-metric obeys the
relation: $<v,v,v,v>_{BM} =v^1 v^2 v^3v^4$.  The  Finslerian
4-metric with the components (\ref{2.11})  coincides up to a
factor with the $n$-dimensional  analogue  of the Berwald-Moor
4-metric. (This metric is a special case of 4th order Shimada
metric \cite{Shimada}.)

\subsection{The equations of motion }

The equations of motion corresponding to the action (\ref{1.1})
have the form
\begin{equation}
{\cal E}_{MN} = \alpha_1 {\cal E}^{(1)}_{MN} + \alpha_2 {\cal
E}^{(2)}_{MN} = 0, \label{1.3e}
\end{equation}
where
\begin{eqnarray}
{\cal E}^{(1)}_{MN} = R_{MN} - \frac{1}{2} R g_{MN},
\label{1.3a} \\
{\cal E}^{(2)}_{MN} = 2(R_{MPQS}R_N^{\ \ PQS} - 2 R_{MP} R_N^{\ \
P}   \nonumber \\
-2 R_{MPNQ} R^{PQ} + R R_{MN}) -  \frac{1}{2} {\cal L}_2  g_{MN}.
\label{1.3b}
\end{eqnarray}

The field equations (\ref{1.3e}) for the metric (\ref{2.2}) are
equivalent to the Lagrange equations   corresponding to the
Lagrangian $L$ from (\ref{2.4})  \cite{Deruelle}.

Thus eqs. (\ref{1.3e}) read
\begin{eqnarray}
\alpha_1  G_{ij} \dot{\beta}^i \dot{\beta}^j - \alpha_2  e^{- 2
\gamma}  G_{ijkl} \dot{\beta}^i \dot{\beta}^j \dot{\beta}^k
\dot{\beta}^l = 0,  \label{2.17} \\
\frac{d}{dt}[  2 \alpha_1  G_{ij} e^{-\gamma + \gamma_0}
\dot{\beta}^j  -  \frac{4}{3} \alpha_2 e^{- 3 \gamma + \gamma_0}
G_{ijkl}  \dot{\beta}^j \dot{\beta}^k \dot{\beta}^l] - L = 0,
\label{2.18}
\end{eqnarray}
$i = 1,\ldots, n$. Due to (\ref{2.17}),
\begin{equation}
L =   \frac{2}{3} e^{-\gamma + \gamma_0} \alpha_1  G_{ij}
\dot{\beta}^i  \dot{\beta}^j. \label{2.18a}
\end{equation}

\section{Exact solutions in  ``pure'' Gauss-Bonnet  model}

Now we put $\alpha_1 = 0$ and $\alpha_2 \neq 0$, i.e. we consider
the cosmological model governed by the action
\begin{equation}
S_2 =  \alpha_2 \int_{M} d^{D}z \sqrt{|g|} {\cal L}_2[g].
\label{3.1}
\end{equation}

The equations of motion (\ref{1.3e}) in this  case read
\begin{equation}
{\cal E}^{(2)}_{MN} = {\cal R}^{(2)}_{MN} - \frac{1}{2} {\cal L}_2
g_{MN} = 0,  \label{3.1a}
\end{equation}
where
\begin{eqnarray}
{\cal R}^{(2)}_{MN} = 2(R_{MPQS}R_N^{\ \ PQS} - 2 R_{MP} R_N^{\ \
P}
\nonumber \\
-2 R_{MPNQ} R^{PQ} + R R_{MN}). \label{3.1b}
\end{eqnarray}

Due to the identity $g^{MN} {\cal R}^{(2)}_{MN} = 2 {\cal L}_2$
the set of eqs.  (\ref{3.1a}) for $D \neq 4$   implies
\begin{equation}
{\cal L}_2  = 0. \label{3.1d}
\end{equation}
It is obvious  that the set of equations  (\ref{3.1a})  is
equivalent for $D \neq 4$  to the  set of equations
\begin{equation}
{\cal R}^{(2)}_{MN} = 0. \label{3.1c}
\end{equation}

The equations of motion (\ref{2.17}), (\ref{2.18})  in this case
read
\begin{eqnarray}
G_{ijkl} \dot{\beta}^i \dot{\beta}^j \dot{\beta}^k
\dot{\beta}^l = 0,  \label{3.2} \\
\frac{d}{dt} \left[ e^{- 3 \gamma + \gamma_0} G_{ijkl}
\dot{\beta}^j \dot{\beta}^k \dot{\beta}^l \right] = 0,
\label{3.3}
\end{eqnarray}
$i = 1,\ldots, n$. Here  $L = 0$ due to  (\ref{3.2}).

Let us put $\ddot{\beta}^i = 0$ for all $i$, or  equivalently,
\begin{equation}
\beta^i = c^i t + c^i_0, \label{3.4}
\end{equation}
where $c^i$ and $c^i_0$ are constants, $i = 1,\ldots, n$.

We also put
\begin{equation}
3 \gamma = \gamma_0 = \sum_{i =1}^{n} \beta^i, \label{3.5}
\end{equation}
i.e. a modified ``harmonic'' time variable is used. Recall that in
the case $\alpha_1 \neq 0$ and $\alpha_2 = 0$, the choice $\gamma
= \gamma_0$ corresponds to the harmonic time variable $t$
\cite{IMZ}.

Then, eqs. (\ref{3.3}) are satisfied identically  and eq.
(\ref{3.2}) gives us the following  constraint
\begin{equation}
G_{ijkl} c^i c^j c^k c^l = 24 \sum_{i < j < k < l} c^i c^j c^k c^l
= 0   \label{3.6}.
\end{equation}

Thus we have obtained an  exact cosmological solution for the
Gauss-Bonnet model (\ref{3.1}), given by the metric (\ref{2.2})
with the functions $\beta^i(t)$ and $\gamma(t)$ from (\ref{3.4})
and  (\ref{3.5}), respectively, and the integration constants
$c^i$ obeying (\ref{3.6}).

\subsection{Power-law solutions}

Let us consider the solutions with $\sum_{i=1}^n c^i \neq 0$.

Introducing the synchronous time variable   $\tau = \frac{1}{c}
\exp(c t + c_0)$,  where   $ c = \frac{1}{3} \sum_{i=1}^n c^i$, $
c_0 = \frac{1}{3} \sum_{i=1}^n  c^i_0$, and defining the  new
parameters  $ p^i =  c^i/c$,  $A_i = \exp[c^i_0 + p^i (\ln c -
c_0)]$,  we get the power-law solution  with the metric
\begin{equation}
g=  - d \tau \otimes d \tau  + \sum_{i=1}^{n}  A_i^2 \tau^{2p^i}
dy^i \otimes dy^i, \label{3.12}
\end{equation}
where  $A_i > 0$ are arbitrary constants, and the parameters
$p^i$ obey the relations
\begin{eqnarray}
\sum_{i=1}^n  p^i = 3,  \label{3.13}\\
G_{ijkl} p^i p^j p^k p^l = 24 \sum_{i < j < k < l} p^i p^j p^k p^l
= 0.  \label{3.14}
\end{eqnarray}

This solution is singular  for any set of parameters
\cite{I-GB-09}. For  $n = 4,5$  it was obtained in
\cite{Deruelle}.

{\bf Example 1.} Let  $D = 6$ and $p_i \neq 0$, $i =1, \dots, 5$.
Then the relations (\ref{3.13}) and (\ref{3.14}) read
\begin{eqnarray}
p^1 + p^2 + p^3 + p^4 + p^5 = 3,  \label{3.13e}\\
p^1 p^2 p^3 p^4 p^5 \left(\frac{1}{p^1}  + \frac{1}{p^2} +
\frac{1}{p^3}  + \frac{1}{p^4} + \frac{1}{p^5} \right)  = 0.
\label{3.14e}
\end{eqnarray}
Let us  put $p^1  = x > 0$, $p^2 = \frac{1}{x}$, $p^3  = z > 0$,
$p^4  = y < 0$, $p^5 = \frac{1}{y}$. Then we get
\begin{equation}
x +  \frac{1}{x} + z + y + \frac{1}{y} = 3, \quad x +  \frac{1}{x}
+ \frac{1}{z} + y + \frac{1}{y} = 0.  \label{3.14ee}
\end{equation}
Subtracting the second relation in (\ref{3.14ee}) from the first
one, we obtain  $z - \frac{1}{z} = 3$ or $z = \frac{1}{2}(3 +
\sqrt{13})$ ($z > 0$).  For any $x > 0$ there are two solutions $y
= y_{\pm}(x)  = \frac{1}{2}(- A \pm \sqrt{A^2 -4})$,  where $A = x
+  \frac{1}{x} + \frac{1}{z} > 2$.

{\bf Proposition.} {\it For $D > 4$, the metric (\ref{3.12}) is  a
solution to  eqs. of motion (\ref{3.1a}) if and only if the set of
parameters $p = (p^1,...,p^n)$ either obeys the relations
(\ref{3.13}) and (\ref{3.14}), or $p = (a,b,0,...,0),
(a,0,b,0,...,0), \ldots $, where $a,b$ are arbitrary real numbers.
}

This proposition is proved in  Appendix C (for $D = 5, 6$ see also
\cite{Deruelle}).

For $D = 2,3,4$ the metric (\ref{3.12}) gives a solution to the
equations of motion (\ref{3.1a}) for any set of parameters $p^i$.

\subsection{Exponential solutions}

Now we consider solutions with  $\sum_{i=1}^n  c^i  = 0$.
Introducing the synchronous time variable   $\tau = t \exp(c_0)$
where $ c_0 = \frac{1}{3} \sum_{i=1}^n  c^i_0$     and defining
the new parameters  $v^i =  c^i \exp(-c_0)$,    $B_i =
\exp(c^i_0)$, we  get a non-singular cosmological  solution  with
the metric
\begin{equation}
g= - d \tau \otimes d \tau  + \sum_{i=1}^{n}  B_i^2 e^{2v^i \tau}
dy^i \otimes dy^i, \label{3.19}
\end{equation}
where  $B_i > 0$ are arbitrary constants, and parameters $v^i$
obey the relations
\begin{eqnarray}
\sum_{i=1}^n  v^i = 0,  \label{3.20}\\
G_{ijkl} v^i v^j v^k v^l = 24 \sum_{i < j < k < l} v^i v^j v^k v^l
= 0.   \label{3.21}
\end{eqnarray}

{\bf Example 2.} Let  $D = 6$ and  $v_i \neq 0$, $i =1, \dots, 5$.
The relations (\ref{3.20}) and (\ref{3.21}) read in this case
\begin{eqnarray}
v^1 + v^2 + v^3 + v^4 + v^5 = 0,  \label{3.20e}\\
v^1 v^2 v^3 v^4 v^5 \left(\frac{1}{v^1}  + \frac{1}{v^2} +
\frac{1}{v^3} + \frac{1}{v^4} + \frac{1}{v^5} \right)  = 0.
\label{3.21e}
\end{eqnarray}
We   put $v^1  = x > 0$, $v^2 = \frac{1}{x}$, $v^3  = 1$, $v^4  =
y < 0$, $v^5 = \frac{1}{y}$. Then we get
\begin{equation}
x +  \frac{1}{x} + 1 + y + \frac{1}{y} = 0,
\label{3.20ee}
\end{equation}
For any $x > 0$ there are two solutions of (\ref{3.20ee}): $y =
y_{\pm}(x)  = \frac{1}{2}(- B \pm \sqrt{B^2 -4})$, where $B = x +
\frac{1}{x} + 1 \geq 3$.

{\bf Remark.} For $D = 4$, or $n= 3$, the equations of motion
(\ref{3.2}), (\ref{3.3}) are satisfied identically for arbitrary
(smooth) functions $\beta^i(t)$ and $\gamma(t)$. This is in an
agreement with the fact that in dimension $D = 4$ the action
(\ref{3.1}) is a topological invariant, and its  variation
vanishes identically.

\section{Reduction to an autonomous set
         of first-order differential equations}

Now we put $\gamma = 0$, i.e. the ``synchronous'' time gauge is
considered. We denote $t = \tau$.  By introducing the
``Hubble-like'' variables $h^i = \dot{\beta}^i$  we rewrite eqs.
 (\ref{2.17}) and (\ref{2.18}) in the following form:

\begin{eqnarray}
\alpha_1  G_{ij} h^i h^j
 - \alpha_2   G_{ijkl} h^i h^j h^k h^l = 0,  \label{5.1} \\
\left[ 2   \alpha_1  G_{ij} h^j -  \frac{4}{3} \alpha_2  G_{ijkl}
h^j h^k h^l \right] \sum_{s=1}^n h^s
\qquad \nonumber \\
 + \frac{d}{d\tau} \left[ 2   \alpha_1  G_{ij} h^j -  \frac{4}{3}
\alpha_2  G_{ijkl}  h^j h^k h^l \right] - L    = 0,   \label{5.2}
\end{eqnarray}
$i = 1,\ldots, n$, where
\begin{equation}
L =   \alpha_1 G_{ij} h^i h^j - \frac{1}{3} \alpha_2   G_{ijkl}
h^i h^j h^k h^l, \label{5.1a}
\end{equation}
see also \cite{Deruelle}.

Due to (\ref{5.1}), $ L =   \frac{2}{3}  \alpha_1  G_{ij} h^i
h^j$.  Thus we obtain an autonomous set of first order
differential equations with respect to  $h^1(\tau), ...,
h^n(\tau)$.

Here we may use the  relations (\ref{2.12}), (\ref{2.13}) and the
following formulae (with $v^i = h^i$)
\begin{eqnarray}
G_{ij}v^j = v^i - S_1,
\label{5.3}   \\
G_{ijkl} v^j v^k v^l = S_1^3  + 2 S_3 -3 S_1 S_2  +  3 (S_2  -
S_1^2)  v^i +  6 S_1 (v^i)^2 - 6(v^i)^3, \label{5.4}
\end{eqnarray}
$i = 1,\ldots, n$, where $S_k = S_k (v) = \sum_{i =1}^n (v^i)^k$.

Let us consider  a fixed point of the system (\ref{5.1}),
(\ref{5.2}):   $h^i(\tau) = v^i$, where the constant vector $v
=(v^i)$  corresponds to the solution
\begin{equation}
\beta^i = v^i \tau +  \beta^i_0,  \label{5.4a}
\end{equation}
$\beta^i_0$ are constants, and $i = 1,\ldots, n$. In this case we
obtain the metric (\ref{3.19}) with the exponential dependence of
the scale  factors. (Another solution   with   $h^i(\tau) =
p^i/\tau$ was obtained earlier in  subsection 3.2.)

Now we put $\alpha_1 \neq 0$ and $\alpha_2 \neq 0$. For fixed the
point $v = (v^i)$ we have the set of polynomial equations
\begin{eqnarray}
G_{ij} v^i v^j - \alpha   G_{ijkl} v^i v^j v^k v^l = 0,
\label{5.5} \\
\left[ 2   G_{ij} v^j
 - \frac{4}{3} \alpha  G_{ijkl}  v^j v^k v^l \right]
\sum_{s=1}^n  v^s - \frac{2}{3} G_{ij} v^i v^j  = 0, \label{5.6}
\end{eqnarray}
$i = 1,\ldots, n$, where  $\alpha = \alpha_2/\alpha_1$. For $n >
3$ it is a set of forth-order polynomial equations.

The trivial solution $v = (v^i) = (0, ..., 0)$ corresponds to  a
flat metric $g$.

For any nontrivial solution $v$ we have $\sum_{i=1}^n v^i \neq 0$
(otherwise one gets from (\ref{5.6}) $G_{ij} v^i v^j = \sum_{i
=1}^{n} (v^i)^2 - (\sum_{i =1}^{n} v^i)^2 = 0$ and hence $v = (0,
\dots, 0)$).

Let us consider the isotropic case $v^1 = ... = v^n = a$. The set
of equations (\ref{5.5}), (\ref{5.6})  is equivalent to the
equation
\begin{equation}
 n(n -1)a^2 + \alpha n(n -1)(n -2)(n -3) a^4 = 0.
 \label{5.7}
\end{equation}
For $n = 1$,  $a$ is arbitrary, and $a =0$ for $n = 2,3$. If $n
> 3$  the nonzero solution to eq. (\ref{5.7}) exists only if
$\alpha  < 0$, and in this case
\begin{equation}
 a = \pm \frac{1}{\sqrt{|\alpha| (n -2)(n -3)}}.
\label{5.8}
\end{equation}

Here arises the problem of classification of all solutions to eqs.
(\ref{5.5}), (\ref{5.6}) for given $n$ . Some special solutions of
the form $(a,...,a,b,...,b)$, e.g., in the context of cosmology
with two factor spaces, for certain dimensions were considered in
literature, see, e.g.,
\cite{Ishihara,ElMakObOsFil,BambaGuoOhta,KirMak}.

Let us outline three  properties of the solutions to the set of
polynomial equations (\ref{5.5}), (\ref{5.6}):

{\bf i)} For any solution $v = (v^1,...,v^n)$, the  vector $(-v) =
(-v^1,...,-v^n)$ is also a solution;

{\bf ii)} For any solution  $v = (v^1,...,v^n)$ and for any
permutation  $\sigma$ of the set of indices  $\{1,..., n \}$, the
vector $v = (v^{\sigma(1)},...,v^{\sigma(n)})$ is also a solution;

{\bf iii)} For any nontrivial solution $v = (v^1,...,v^n) \neq
(0,...,0)$  there are no more than  three different numbers among
$v^1,...,v^n$.

The first proposition is trivial. The second one simply follows
from the relations (\ref{2.12}), (\ref{2.13}), (\ref{5.3}),
(\ref{5.4}).

Let us prove the  third proposition. Suppose that there exists a
nontrivial solution $v = (v^1,...,v^n)$ with more than three
different  numbers among  $v^1,...,v^n$. Due to (\ref{5.4}),
(\ref{5.6}) and  $\sum_{i=1}^n v^i  \neq 0$, any number $v^i$
obeys the cubic equation $C_0 + C_1 v^i + C_2 (v^i)^2  +
C_3(v^i)^3 = 0$, with $C_3 \neq 0$,  $i = 1,\ldots, n$, and hence
at most three numbers among $v^i$  may be different. Thus we
obtain a contradiction. The proposition {\bf iii)} is proved.

This implies that in future investigations of solutions to eqs.
(\ref{5.5}), (\ref{5.6}) for arbitrary $n$ we will to consider
three nontrivial cases such that: 1) $v = (a,...,a)$ (see
(\ref{5.8})); 2) $v = (a,...,a,b,...,b)$  ($a \neq b$); and 3) $v
= (a,...,a,b,...,b,c,...,c)$ ($a \neq b$, $b \neq c$, $a \neq c$).
One may also put   $a > 0$ due to item {\bf i)}.

%%%%%%%%%%%%%%%%%%%%%%%%%%%%%%%%%%%%%%%%%%%%%%%%%%%%%%%%%%%%%%%%%%%%%%%%%
\section{Conclusions and discussions}
%%%%%%%%%%%%%%%%%%%%%%%%%%%%%%%%%%%%%%%%%%%%%%%%%%%%%%%%%%%%%%%%%%%%%%%%%

We have considered the  $(n +1)$-dimensional Einstein-Gauss-Bonnet
model.  For  diagonal  cosmological metrics, we have written  the
equations of motion as a set of Lagrange equations  (see also
\cite{Deruelle})  with   the effective Lagrangian  governed by two
``minisuperspace'' metrics on $\R^{n}$: (i) the pseudo-Euclidean
2-metric (corresponding to the scalar curvature term)  and  (ii)
the Finslerian  4-metric (corresponding to the Gauss-Bonnet term).
The  Finslerian 4-metric is proportional to the $n$-dimensional
Berwald-Moor 4-metric (it is a special case of the  Shimada
quartic metric \cite{Shimada}). Thus we have found rather a
natural and ``legitimate'' application of the $n$-dimensional
Berwald-Moor  metric ($n = 4, 5, \dots$) in $(n+1)$-dimensional
gravity  with a Gauss-Bonnet term. The effective Lagrangian
(\ref{2.4}) was considered  earlier by N. Deruelle in
\cite{Deruelle} (for $\gamma = 0$). (See also \cite{Pavl} and
references therein.)  Here we put an additional accent on  the
Finslerian (Berwald-Moor) structure of the second term in
(\ref{2.4}).

For the case of the pure Gauss-Bonnet model,   we have derived two
exact solutions  with power-law and exponential   dependences of
the scale factors  on the synchronous time variable. The first
(power-law) solution was obtained earlier by N. Deruelle for $n =
4, 5$ \cite{Deruelle}  and verified by A. Toporensky and P.
Tretyakov   for $n = 6,7$ \cite{TT}. In  \cite{Pavl} this solution
was  verified for all $n$.

When  the synchronous time  gauge was considered, the equations of
motion were reduced to  an autonomous set of first-order
differential equations (see also \cite{Deruelle}). For $\alpha_1
\neq 0$ and $\alpha_2 \neq 0$  it was shown that for any
nontrivial  solution with the exponential time dependence of the
scale factors $a_i(\tau) = A_i \exp( v^i  \tau)$, there are no
more than three different  numbers among  $v^1,...,v^n$. This
means that  solutions of this type have a restricted anisotropy.
Such solutions  may be used for construction  of new cosmological
solutions, e.g., describing an accelerated expansion of our
3-dimensional factor space  and small enough variation of the
effective gravitational constant. For this approach see
\cite{BZhuk,IKM-08} and references therein.

Here an open problem arises: do the solutions (\ref{1.3}),
(\ref{1.7})   with ``jumping''  parameters $p^i, A_i$ appear as
asymptotic  solutions in EGB  gravity (for some $n$)  when
approaching  a singular  point?   Recall that the Kasner-type
solutions with ``jumping''   parameters $p^i, A_i$ describe an
approach to a singular  point  in certain gravitational models,
e.g., with   matter sources, see
\cite{BLK,DHSp,IKM-bil,IM-bil,DamH1,DHN,IM-bil-rev} and references
therein. This problem may be a subject of   separate
investigations.   [Worth mentioning is the paper by T. Damour and
H. Nicolai \cite{DamNic}, which  includes a study of the effect of
the 4th order (in curvature)  gravity terms, including the
Euler-Lovelock term (octic in  velocities), and its compatibility
with the Kac-Moody algebra  $E_{10}$.]

Other applications of  the Lagrange approach considered in
\cite{Deruelle}  and  in the present paper  will be connected with
inclusion of scalar fields  and a generalization to the Lowelock
model \cite{Low}  (see also \cite{Deruelle,Pavl}).

\begin{center}
{\bf Acknowledgments}
\end{center}

This work was supported in part by the Russian Foundation for
Basic Research grant No. 09-02-00677-a. The author is also
grateful to A.V. Toporensky and D.G. Pavlov for stimulating
lectures  and discussions.

\renewcommand{\theequation}{\Alph{subsection}.\arabic{equation}}
\renewcommand{\thesection}{}
\renewcommand{\thesubsection}{\Alph{subsection}}
\setcounter{section}{0}

\section{Appendix}

\subsection{$(1+n)$-splitting}

Consider the metric defined on $\R_{*} \times \R^{n}$ ($\R_{*} =
(t_{-},t_{+})$ is an open subset in $\R$)
\begin{equation}
g= - e^{2{\gamma}(t)} dt \otimes dt  + \sum_{i,j =1}^{n} h_{ij}(t)
dy^i \otimes dy^j. \label{A.1}
\end{equation}

Here $(h_{ij}(t))$ is a symmetric nondegenerate matrix for any $t
\in \R_{*}$, smoothly depending on $u$.  The function
${\gamma}(t)$ is smooth.

Calculations give the following nonvanishing (identically)
components of the Riemann tensor
\begin{eqnarray}
R_{0i0j} = - R_{i00j} = - R_{0ij0} = R_{i0j0} = \frac{1}{4} [-2
\ddot{h}_{ij} + 2 \dot{\gamma} \dot{h}_{ij} +
\dot{h}_{ik} h^{kl} \dot{h}_{lj}]  ,  \label{A.2}\\
R_{ijkl} =  \frac{1}{4}  e^{- 2 \gamma} (\dot{h}_{ik} \dot{h}_{jl}
- \dot{h}_{il} \dot{h}_{jk}),  \label{A.3}
\end{eqnarray}
$i,j,k,l = 1, \dots, n$, where $h^{-1} = (h^{ij})$ is the matrix
inverse to the matrix $h = (h_{ij})$. We denote $\dot{A} = dA/dt$
etc.

For nonzero (identically)   components of the Ricci tensor we get
\begin{eqnarray}
R_{00} =  \frac{1}{2} [- h^{il} \ddot{h}_{li} + \frac{1}{2} h^{ij}
\dot{h}_{jk}  h^{kl} \dot{h}_{li} +
h^{ik} \dot{h}_{ki}  \dot{\gamma}],  \label{A.4}\\
R_{ij} =  \frac{1}{4}  e^{- 2 \gamma} [ 2 \ddot{h}_{ij} +
\dot{h}_{ij}(h^{kl} \dot{h}_{lk} - 2  \dot{\gamma}) -2
\dot{h}_{ik} h^{kl} \dot{h}_{lj} ], \label{A.5}
\end{eqnarray}
$i,j = 1, \dots, n$.

The scalar curvature reads
\begin{equation}
R =  \frac{1}{4} e^{- 2 \gamma} [ 4 {\rm tr}(\ddot{h}h^{-1}) +
{\rm tr}(\dot{h}h^{-1}) ({\rm tr}(\dot{h}h^{-1}) - 4 \dot{\gamma})
- 3  {\rm tr}(\dot{h} h^{-1}\dot{h}h^{-1})].   \label{A.6}
\end{equation}

\addtocounter{section}{1} \setcounter{equation}{0}
\subsection{$f$-function}

The  function $f$ in (\ref{2.3}) has the following form (see
\cite{Deruelle} for $\gamma = 0$)
\begin{equation}
f = \alpha_1 f_1 +  \alpha_2 f_2, \label{B.1}
\end{equation}
where
\begin{eqnarray}
f_1 = 2  e^{-\gamma + \gamma_0} \sum_{i =1}^{n} \dot{\beta}^i,
\label{B.2}   \\
f_2  = \frac{4}{3}  e^{- 3 \gamma + \gamma_0} \left[ 2 \sum_{i
=1}^{n} (\dot{\beta}^i)^3  - 3 (\sum_{i =1}^{n} \dot{\beta}^i)
\sum_{j =1}^{n} (\dot{\beta}^i)^2 + ( \sum_{i =1}^{n}
\dot{\beta}^i)^3 \right]. \label{B.3}
\end{eqnarray}

The function $f_2$ may be rewritten as follows:
\begin{equation}
f_2  = \frac{4}{3}  e^{- 3 \gamma + \gamma_0} G_{ijk}
\dot{\beta}^i \dot{\beta}^j \dot{\beta}^k, \label{B.4}
\end{equation}
where
\begin{equation}
G_{ijk}  = (\delta_{ij} -1)(\delta_{ik} -1)(\delta_{jk} -1)
\label{B.5}
\end{equation}
are  components of  a  Finslerian 3-metric.

\addtocounter{section}{1} \setcounter{equation}{0}
\subsection{Proof of the Proposition in Sec. 3.1}

The equations of motion  (\ref{5.1}) and (\ref{5.2}) corresponding
to the metric  (\ref{3.12}) with $h^i = p^i/\tau$  (here $\alpha_1
=0$  and $\alpha_2 \neq 0$)  read
\begin{eqnarray}
{\cal A} \equiv  G_{ijkl}p^i p^j p^k p^l = 0,
\label{E.1}     \\
{\cal D}_i \equiv G_{ijkl} p^j p^k p^l  = 0, \label{E.2}
\end{eqnarray}
$i = 1,\dots, n$.

Let $D = n+1 \neq 4$  and
\begin{equation}
{\cal B} \equiv  \frac{1}{(n - 3)} \sum_{i =1}^{n} {\cal D}_i,
\label{E.6}
\qquad {\cal C}_i \equiv \frac{1}{3} ({\cal B} - {\cal
D}_i),
\end{equation}
$i = 1,\dots, n$.

For $D \neq 4$ the set of  eqs. (\ref{E.1}), (\ref{E.2}) is
equivalent to the following set of equations:
\begin{eqnarray}
{\cal A} =  S_1^4 -  6  S_1^2 S_2 + 3 S_2^2 + 8  S_1 S_3 - 6 S_4 =
24 \sum_{i < j < k < l} p^i p^j p^k p^l = 0,
\label{E.10}     \\
{\cal B} =  (S_1 - 3)(S_1^3  - 3 S_1 S_2 + 2 S_3) = 6(S_1 -
3)\sum_{i < j < k } p^i p^j p^k = 0,
\label{E.11} \\
{\cal C}_i =  (S_1 - 3) p^i [2 (p^i)^2 - 2 S_1 p^i + S_1^2 - S_2 ]
= 0,
\label{E.12}
\end{eqnarray}
$i = 1,\dots, n$. Here $S_k = S_k (p) = \sum_{i =1}^n (p^i)^k$,
and we  have used the identities (\ref{2.13}), (\ref{5.4}) and the
following identity:
\begin{equation}
S_1^3  - 3 S_1 S_2 + 2 S_3 = G_{ijk}p^i p^j p^k = 6 \sum_{i < j <
k} p^i p^j  p^k,
\label{E.8}
\end{equation}
where $G_{ijk}$ are defined in (\ref{B.5}).

For $S_1 = 3$ we obtain the main solution governed by the
relations (\ref{3.13}) and (\ref{3.14}).

Now consider another case, $S_1 \neq 3$. Let $k$ be the number of
all nonzero numbers  among $p^1,...,p^n$. For $k = 0$ we get the
trivial solution $(0,...,0)$.  Let $k \geq 1$. We suppose without
loss of generality that   $p^1,...,p^k$ are nonzero. For $k = 1,
2$ all relations  (\ref{E.10})-(\ref{E.12}) are satisfied
identically.  In all three cases $k = 0, 1, 2$  the solutions have
the form $(a,b,0..,0)$ (plus permutations for the general setup).

Consider $k \geq 3$. From (\ref{E.12}) and $S_1 \neq 3$ we obtain
\begin{equation}
2 (p^i)^2 - 2 S_1 p^i + S_1^2 - S_2 = 0,
\label{E.13}
\end{equation}
$i = 1,\dots, k$. Summing  over $i$ gives us $(2 -k) (S_2 - S_1^2)
= 0$, or $S_2 = S_1^2$. Then we obtain from (\ref{E.11}) $S_3 =
S_1^3$ and from (\ref{E.10}): $S_4 = S_1^4$. Thus we get $S_4 =
S_2^2$ implying $\Sigma = \sum_{1 \leq i < j \leq k } (p^i)^2
(p^j)^2 =  0$. But  $\Sigma \geq (p^1)^2 (p^2)^2 > 0$. Thus we
obtain a contradiction. That means that for $S_1 \neq 3$ we have
only solutions with $k \leq 2$  of the form $(a,b,0..,0)$ (plus
permutations for general setup). The Proposition in Subsection 3.1
is proved.

\end{document}